\documentstyle[11pt]{article}
\begin{document}
\baselineskip 14pt
\begin{titlepage}
Preprint-NIM-TPD-001/(1998)

\vspace{1cm}
\centerline{\large\bf Shift Operators for XXX-Heisenberg Chain }
\vspace{1cm}

\centerline{ Jing-Ling Chen$^1$, Mo-Lin Ge$^{1,2}$, and Kang Xue$^{2,3}$}
\vspace{0.5cm}
{\small
\centerline{\bf 1. Theoretical Physics Division,}
\centerline{\bf Nankai Institute of Mathematics,}
\centerline{\bf Nankai University, Tianjin 300071, P.R.China}
\centerline{ Email:   yongliu@sun.nankai.edu.cn}
\centerline{ Fax: 0086-22-23501532, Tel: 0086-22-23501029}
\vspace{0.2cm}
\centerline{\bf 2. Center for Advanced Study }
\centerline{\bf Tsinghua University, Beijing 100084, P.R.China}
\vspace{0.2cm}
\centerline{\bf 3. Physics Department, Northeast Normal University,}
\centerline{\bf Changchun,Jilin, 130024, P.R.China}
}
\vspace{2cm}

\centerline{\bf Abstract}

The shift operators for XXX-Heisenberg chain are found. They are formed by
local Yangian operators and the amplitutes of the eigenfunctions obeying
Bethe ansatz for the Hamiltonian. The physical implication of the shift
operators are also explained.

\vspace{2cm}



\end{titlepage}

\centerline{\large\bf Shift Operators for XXX-Heisenberg Chain }
\vspace{1cm}
{\small

The XXX-Heisenberg chain(HC) defined by
\begin{equation}
H=J\sum_{j=1}^{N}({\vec {\bf S}}_j\cdot{\vec {\bf S}}_{j+1}-\frac{1}{4}),
\end{equation}
is exactly solved by the Bethe ansatz \cite{bethe}. The model describes
nearest-neighbour interacting spins, situated on the sites of a periodic
lattice $N$. For $J>0$ and $J<0$, it corresponds to the anti-ferromagnetic
and ferromagnetic spin chain, respectively. The simple form of equation
(1) bellies the rich physical behaviour that it displays, and an
understanding of the physics of the HC in one-dimension has proved a
formidable task for theoretical and methematical physicists over the last
six decades \cite{faddeev} \cite{schulz} \cite{affleck} \cite{anderson}
\cite{batchelor} \cite{shastry} \cite{essler} \cite{tennant} \cite{grabowski}
\cite{hosotani}. Explict form of the
eigenfunctions with $r(r=0,1,2,\cdots,N)$ down-spins (i.e. with the
eigenvalue $S_z=\frac{N}{2}-r$) are given by\cite{bethe}
\begin{equation}
\label{eqpsir}
\mid \psi_r>=\sum_{m_1<m_2<\cdots<m_r}
a(m_1,m_2,\cdots,m_r)\phi( m_1,m_2,\cdots,m_r),
\end{equation}
where $\phi (m_1,m_2,\cdots,m_r)$ represents a spin state with $r$ down-spins
on the $m_j$-th $(j=1,2,\cdots,r)$ sites. The coefficients
\begin{equation}
a(m_1,m_2,\cdots,m_r)=\sum_{P=1}^{r!}\exp[i(\sum_{k=1}^{r}
{\theta_{P_k}}m_k+\frac{1}{2}\sum\phi_{P_k,P_n})],
\end{equation}
are defined only for the ordering $m_1<m_2<\cdots<m_r$, and $P$ is any
permutation of the $r$ numbers $1,2,\cdots, r$,  $P_k$ the number replaced
$k$ under this permutation, and $\phi_{kn}=-\phi_{nk}$. In
eq.(\ref{eqpsir}) for the case with $r=0$
$$
   \mid \psi_0>=\mid \uparrow \uparrow \cdots \uparrow > ,
$$
is the vaccum state with all spins up. In the Bethe ansatz eigenfunctions
$\mid \psi_r>$ $(r=1,2,\cdot,N)$ are always generated from $\mid \psi_0>$
by the operator
\begin{equation}
\label{eqqr0}
[\; {\cal Q}_{r}^+ = \sum_{m_1<m_2<\cdots <m_r }a(m_1,m_2,\cdots, m_r )
S_{m_1}^- S_{m_2}^- \cdots S_{m_r}^-\;]\; \mid \psi_0>=\mid \psi_r> ,
\end{equation}
that has been well known. On the other hand, from the point of view of the
operator method, if shift operators of a Hamiltonian (or other observables)
known, then the eigen-problem can be solved exactly. Inversly, if a
Hamiltonian system can be solved exactly, in principle, it must have shift
operators. Shift operator transforming neighboured level states are discussed
extensively in the linear systems in quantum mechanics, such as the harmonic
oscillator and the hydrogen atom \cite{lange}. It is natural to ask whether
such kind of operators(not the cases in eq.(\ref{eqqr0})) do exist in the
non-linear models, such as XXX-Heisenberg chain. The purpose of this paper is
just to establish shift operators transform $\mid \psi_{r-1}>$ to
$\mid \psi_r>$ in the model.

We start from introducing the unified shift operators
\begin{equation}
\label{eqqr1}
 Q_{r,r-1}^+ = -i \sum_{j<k}^{N} W_{jk}^{(r)}({\vec S}_j \times {\vec S}_k)^-,
\;\; W_{jk}^{(r)}=-W_{kj}^{(r)},
\end{equation}
in $\mid \psi_{r-1}>$ there are already $(r-1)$ down-spins, i.e. acting
$-i ({\vec S}_j \times {\vec S}_k)^- = S_j^- S_k^z - S_k^- S_j^z$  on
$\mid \psi_{r-1}>$ will lead to a state with $r$ down-spins:
\begin{equation}
\label{eqqr2}
 Q_{r,r-1}^+
 \; \mid \psi_{r-1}>=\mid \psi_r>,
\end{equation}
namely, $Q_{r,r-1}^+$ transforms the adjacent eigenfunctions specified by
$r-1$ and $r$. What we need to do is to determine the coefficents
$W_{jk}^{(r)}$ for given states $\mid \psi_{r-1}>$ and $\mid \psi_r>$.
It is worth mentioning that the operator
${\vec {\bf J}}= -i \sum\limits_{j<k}^{N} {\vec S}_j \times {\vec S}_k $
is a Yangian operator. Operator $Q_{r,r-1}^+$ is yielded by combining
$W_{jk}^{(r)}$ with local Yangian operator
${\vec S}_j \times {\vec S}_k$ and summation, hence $Q_{r,r-1}^+$ has
the most natural form and is a generalization of the Yangian operator.

Taking eq.(\ref{eqqr2}) and $H \mid \psi_{r}>=E_r \mid \psi_{r}>$ into
account, one can verify that
\begin{equation}
\label{eqHQ}
(\;[H, Q_{r,r-1}^+] = \omega_{r,r-1} Q_{r,r-1}^+ \;)\; \mid \psi_{r-1}>,
\end{equation}
where
$$
   \omega_{r,r-1} = E_r- E_{r-1} ,
$$
i.e. the energy interval. Therefore in the sense of acting the operators on
eigenfunctions of $H$, $Q_{r,r-1}^+$ are shift operators of $H$.

To find the explict form of $Q_{r,r-1}^+$, we
need to determine $W_{jk}^{(r)}$. We consider the cases $r=1,2,\cdots$
successively.

(a) $r=1$. In this case, eq.(\ref{eqqr2}) becomes
\begin{equation}
\label{eqq0to1}
[\; Q_{1,0}^+ = -i \sum_{j<k}^{N} W_{jk}^{(1)}({\vec S}_j \times {\vec S}_k)^-
 \; \mid \psi_{0}>=\mid \psi_1>.
\end{equation}
Because
$$
   \mid \psi_1>=\sum_{m=1}^{N}a(m)\phi(m),
$$
and
$$
[\; -i ({\vec S}_j \times {\vec S}_k)^- =S_j^- S_k^z - S_k^- S_j^z \;]
\mid \stackrel{j}{\uparrow}
\stackrel{k}{\uparrow} > = \frac{1}{2}
(\mid \stackrel{j}{\downarrow} \stackrel{k}{\uparrow} > -
\mid \stackrel{j}{\uparrow} \stackrel{k}{\downarrow} > ),
$$
by comparing the corresponding coefficients of $\phi(m)$ of the both sides of
eq.({\ref{eqq0to1}}), we have
\begin{equation}
\label{w1}
 \frac{1}{2}[W_{m1}^{(1)}+\cdots+W_{m,m-1}^{(1)}+W_{m,m+1}^{(1)}+
\cdots+W_{mN}^{(1)}] =a(m),
\;\;\; (m=1,2,\cdots).
\end{equation}

Under periodic boundary condition it is well-known:
$$
a(m)=\exp (im\theta), \;\;\;
\sum_{m=1}^{N}a(m)= 0,
$$
with
$$
\theta =\frac{2\pi}{N}n;\ \ n=0,\pm 1,\cdots,\pm (\frac{N}{2}-1),
\pm \frac{N}{2}.
$$
Therefore, one obtain the solutions of eq.(\ref{w1})
\begin{equation}
\label{eqwjk1}
W_{jk}^{(1)}=\frac{2}{N}(\exp (ij\theta)-\exp (ik\theta)),\;\;
(j,k=1,2,\cdots,N).
\end{equation}

There are many ways to transform $\mid \psi_0>$ to $\mid \psi_1>$, however
as we see later, the form of eq.(\ref{eqwjk1}) can be extented to more
general one transforming $\mid \psi_{r-1}>$ and $\mid \psi_r>$.

(b) $r=2$. In this case, eq.(\ref{eqqr2}) becomes
\begin{equation}
\label{eqq1to2}
[\; Q_{2,1}^+ = -i \sum_{j<k}^{N} W_{jk}^{(2)}({\vec S}_j \times {\vec S}_k)^-
 \; \mid \psi_{1}>=\mid \psi_2>,
\end{equation}
and
$$
\mid \psi_2>=\sum_{m_1<m_2}^{N}a(m_1,m_2)\phi(m_1,m_2),\;\;
(m_1,m_2=1,2,\cdots,N)
$$
where
$$
\phi(m_1,m_2)=\mid \cdots \stackrel{m_1}{\downarrow} \cdots
\stackrel{m_2}{\downarrow} \cdots  >
$$
represents spin-down state that on the $m_j$-th $(j=1,2)$ sites. From the
known result of the Bethe Ansatz, one has
$$
a(m,k)=Ce^{im\theta_1}e^{ik\theta_2}+C'e^{im\theta_2}e^{ik\theta_1},\;\;
(m,k=1,2,\cdots,N)
$$
with
$$C=\exp(i\frac{\phi_{12}}{2}),\ \
C'=\exp(-i\frac{\phi_{12}}{2}),$$
and
$$
2\cot{\frac{\phi_{12}}{2}}=\cot \frac{\theta_1}{2}-\cot \frac{\theta_2}{2},
\hspace{1cm} -\pi \le \phi_{12}  \le \pi.
$$
i.e. $C$ and $C'$ depent only on $\theta_1$ and $\theta_2$,  not on
$m_1$ and $m_2$ (or $m$ and $k$ ).

Noting that
\begin{equation}
\label{eqsjk2}
-i ({\vec S}_j \times {\vec S}_k)^- \mid \stackrel{j}{\downarrow}
\stackrel{k}{\uparrow} > = \frac{1}{2}
\mid \stackrel{j}{\downarrow} \stackrel{k}{\downarrow} > ;
\;\;\;
-i ({\vec S}_j \times {\vec S}_k)^- \mid \stackrel{j}{\uparrow}
\stackrel{k}{\downarrow} > = -\frac{1}{2}
\mid \stackrel{j}{\downarrow} \stackrel{k}{\downarrow} > ,
\end{equation}
and by comparing the corresponding coefficients of $\phi(m_1,m_2)$ of the
both sides of eq.({\ref{eqq1to2}}), we have
\begin{equation}
\label{briefw}
\frac{1}{2}W_{m_1,m_2}^{(2)}[a(m_1)-a(m_2)]+
\frac{1}{2}\sum_{m\ne m_1,m_2}^{N}[W_{m_2,m}^{(2)}a(m_1)+
W_{m_1,m}^{(2)}a(m_2)]=a(m_1,m_2).
\end{equation}
with $ m_1,m_2=1,2,\cdots,N $.

Now we come to solve eq.(\ref{briefw}), i.e. to find the solutions of
$W_{m_1,m_2}^{(2)}$.

Because
$$
a(m_1)=\exp (im_1\theta),\;\;\;a(m_2)=\exp (im_2\theta),
$$
we have
$$
\frac{1}{i\theta}\frac{\partial}{\partial m_1} a(m_1)=a(m_1),\;\;
\frac{1}{i\theta}\frac{\partial}{\partial m_1} a(m_2)=0,
 $$
$$
\frac{1}{i\theta}\frac{\partial}{\partial m_1} a(m_2)=a(m_2),\;\;
\frac{1}{i\theta}\frac{\partial}{\partial m_1} a(m_1)=0,
$$
or in general
\begin{equation}
\label{partial2}
\frac{1}{i\theta}\frac{\partial}{\partial m} a(j)=\delta_{jm}a(m),
\;\;\;(m,j=1,2,\cdots,N),
\end{equation}
where $m=x_m$ represents the coordinate of the spin stays on the $m-$th
site of the lattice.

From eq.(\ref{partial2}), one can obtain
\begin{equation}
\label{select1}
\frac{1}{i\theta}\frac{\partial}{\partial m}\mid \psi_1>=
\frac{1}{i\theta}\frac{\partial}{\partial m}\sum_{m=1}^{N}a(m)\phi(m)
=a(m)\phi(m),
\end{equation}
hence, the action of the partial differential operator
$\frac{1}{i\theta}\frac{\partial}{\partial m}$ is clear, when it
acts on $\mid \psi_1 >$, it will pick up the term $a(m)\phi(m)$ among
$\mid \psi_1 >$.
           
From the above analysis, if we set
\begin{equation}
\label{soluwjk}
W_{j,k}^{(2)}=\frac{1}{i\theta}
A(j,k)\;
[ \frac{1}{a(j)} \frac{\partial}{\partial (m=j)}-
  \frac{1}{a(k)} \frac{\partial}{\partial (m=k)} ],
\end{equation}
with
$$
A(j,k)= A(k,j)= \left \{
  \begin{array}{ll}
     a(j,k)   &  if \;\;j<k,  \\
     a(k,j)   &  if \;\;j>k,
  \end{array}
   \right.
$$
then eq.(\ref{briefw}) is satisfied.

One may argue that why $W_{j,k}^{(2)}$ is not a number, but a partial
differential operator acting on the coefficients $a(j)$. If
$W_{j,k}^{(2)}$ is a number, then eq.(\ref{briefw}) can be viewed as a
linear $N(N-1)/2$ set of equations. In the procedure for
determining the coefficients $W_{j,k}^{(2)}$, one finds that he must
set a relation between $a(j)$ and $a(j,k)$,$(j,k=1,2,\cdots,N)$. However,
the relation cannot be valid when one substitutes the results of $a(j)$ and
$a(j,k)$ obtained from the Bethe ansatz. The fact can easily be verified
by taking the example in the case with $N=4$. Therefore, $W_{j,k}^{(2)}$
cannot be numbers, but may be operators.

In the following, we shall derive the shift operator $Q_{2,1}^+$ from
physical consideration, so that the physical picture for the transformation
from $\mid \psi_1>$ to $\mid \psi_2>$ can be seen clearly. In the former
state, one of $N$ spins is down, while in the latter, two of $N$ spins are
down. The crucial point is that, based on $\mid \psi_1>$ in which there has
already been one down-spin, how we can invert the second one so that
it can be shift to be $\mid \psi_2>$. Firstly, we must know clearly on which
site the spin is down. For a given term
$$
a(m)\phi(m)=a(m) \mid \stackrel{1}{\uparrow} \stackrel{2}{\uparrow} \cdots
\stackrel{m}{\downarrow} \cdots \stackrel{N}{\uparrow}  >, $$
it indicates that only on the $m-$th site the spin-state is down, and its
amplitute is $a(m)$. With the help of the partial differential operator
$\frac{1}{i\theta}\frac{\partial}{\partial m}$, the wanted term can be
picked up from $\mid \psi_1>$ (see eq.(\ref{select1})). After this
manipulation, we know clearly that on the $m-$th site the spin has
already been down, thus the second down-spin can occur on the $j-$th
$ (j=1, 2, \cdots, m-1, m+1, \cdots, N)$ sites, respectively.
Meanwhile, $\mid \psi_2 >$ can be rewritten as
$$
\mid \psi_2 >= \frac{1}{2}  \sum_{m=1}^{N} \Phi(m),
$$
\begin{equation}
 \Phi(m)= \sum_{j=1}^{m-1}a(j,m)\phi(j,m) +
          \sum_{k=m+1}^{N}a(m,k)\phi(m,k).
\end{equation}

Introducing
$$
T_m=[ \; \sum_{j=1}^{m-1}T_{m \rightarrow (j,m)}^-  +
     \sum_{k=m+1}^{N}T_{m \rightarrow (m,k)}^- \;],
$$
since
$$
[ \; T_{m \rightarrow (j,m)}^-=
 \frac{a(j,m)}{a(m)}(2i)({\vec S}_{j} \times {\vec S}_{m})^- \;]
a(m)\phi(m)= a(j,m)\phi(j,m),\;\;\;(j<m),
$$
$$
[ \; T_{m \rightarrow (m,k)}^-=
 \frac{a(m,k)}{a(m)}(-2i)({\vec S}_{m} \times {\vec S}_{k})^- \;]
a(m)\phi(m)= a(m,k)\phi(m,k),\;\;\;(k>m),
$$
one obtains
$$
T_m a(m)\phi(m) = \Phi(m).
$$
Define
$$
F_m= T_m \frac{1}{i\theta}\frac{\partial}{\partial m},
$$
we have
$$
F_m \mid \psi_1 >=
        \sum_{j=1}^{m-1}a(j,m)\phi(j,m) +
        \sum_{k=m+1}^{N}a(m,k)\phi(m,k)
$$
so that
\begin{equation}
(\sum_{m=1}^{N}F_m) \mid \psi_1 >= 2 \mid \psi_2 >.
\end{equation}
Since (see eq.(\ref{eqqr1}))
$$
[\; Q_{2,1}^+ = -i \sum_{j<k}^{N} W_{jk}^{(2)}({\vec S}_j \times {\vec S}_k)^-\;]
\mid \psi_1>=\mid \psi_2>, $$
we then arrive at
\begin{equation}
\label{eqqf}
Q_{2,1}^+ = \frac{1}{2}\sum_{m=1}^{N}F_m,
\end{equation}
by identifying the coefficients of $({\vec S}_j \times {\vec S}_k)^-$ of the
both sides of eq.(\ref{eqqf}), it leads to
$$
W_{j,m}^{(2)}=\frac{1}{i\theta}
a(j,m)\;
[ \frac{1}{a(j)} \frac{\partial}{\partial j}-
  \frac{1}{a(m)} \frac{\partial}{\partial m} ],  $$
\begin{equation}
W_{m,k}^{(2)}=\frac{1}{i\theta}
a(m,k)\;
[ \frac{1}{a(m)} \frac{\partial}{\partial m}-
  \frac{1}{a(k)} \frac{\partial}{\partial k} ],
\end{equation}
i.e. they are nothing but the explicit expression of eq.(\ref{soluwjk}).

Therefore, the physical picture for the transformation from $\mid \psi_1>$
to $\mid \psi_2>$ is clear. This idea can be extented to generate
$\mid \psi_{r}>$ for given $\mid \psi_{r-1}>$, for example, in the  similar
manner we can find the shift operator transforming $\mid \psi_2>$ to
$\mid \psi_3>$.

(c) $r=3$.

Similar to (b), the eigenfunction
$$
\mid \psi_3>=\sum_{m_1<m_2<m_3}^{N}a(m_1,m_2,m_3)\phi(m_1,m_2,m_3),\;\;
(m_1,m_2,m_3=1,2,\cdots,N)
$$
can be recast to
$$
\mid \psi_3>= \frac{1}{3} \sum_{m_1<m_2}^{N} \Phi(m_1,m_2),
$$
where
$$
\Phi(m_1,m_2) =
     \sum_{j=1}^{m_1-1} a(j,m_1,m_2)\phi(j,m_1,m_2) +
$$
$$
     \sum_{k=m_1+1}^{m_2-1} a(m_1,k,m_2)\phi(m_1,k,m_2) +
     \sum_{l=m_2+1}^{N} a(m_1,m_2,l)\phi(m_1,m_2,l) .
$$

In $\mid \psi_2>$ two of $N$ spins are down, while in $\mid \psi_3>$ three
are three down-spins. Just like the case in (b), at first, we introduce the
partial differential operator
$\frac{1}{i\theta_1}\frac{1}{i\theta_2}
\frac{\partial}{\partial m_1}\frac{\partial}{\partial m_2}$, which yields
\begin{equation}
\label{selem1m2}
\frac{1}{i\theta_1}\frac{1}{i\theta_2}
\frac{\partial}{\partial m_1}\frac{\partial}{\partial m_2}
\mid \psi_2>=
a(m_1,m_2)\phi(m_1,m_2),
\end{equation}
i.e. the term
$$
a(m_1,m_2)\phi(m_1,m_2) = a(m_1,m_2)
\mid \cdots \stackrel{m_1}{\downarrow} \cdots \stackrel{m_2}{\downarrow}
\cdots  > $$
is picked up from $\mid \psi_2>$, so that two down-spins on the $m_1-$th and
$m_2-$th sites have been pre-set. Further, the third down-spin can occur
on the $j-$th site:
$$
(j=1, 2, \cdots, m_1-1, m_1+1, m_1+2 ,\cdots, m_2-1,
m_2+1, m_2+2, \cdots, N).$$

Introducing
$$
T_{(m_1,m_2)}=
     \sum_{j=1}^{m_1-1} T_{(m_1,m_2) \rightarrow (j,m_1,m_2)}^- +
     \sum_{k=m_1+1}^{m_2-1} T_{(m_1,m_2) \rightarrow (m_1,k,m_2)}^-
     + \sum_{l=m_2+1}^{N} T_{(m_1,m_2) \rightarrow (m_1,m_2,l)}^-,
$$
where
$$
 T_{(m_1,m_2) \rightarrow (j,m_1,m_2)}^-=
 \frac{a(j,m_1,m_2)}{a(m_1,m_2)} \;(2i) [\;
 ({\vec S}_{j} \times {\vec S}_{m_1})^-
 +
 ({\vec S}_{j} \times {\vec S}_{m_2})^- \;];
$$
$$
 T_{(m_1,m_2) \rightarrow (m_1,k,m_2)}^-=
 \frac{a(m_1,k,m_2)}{a(m_1,m_2)} \;(2i) [\;
 -
 ({\vec S}_{m_1} \times {\vec S}_{k})^-
 +
 ({\vec S}_{k} \times {\vec S}_{m_2})^- \;];
$$
and
$$
 T_{(m_1,m_2) \rightarrow (m_1,m_2,l)}^-=
 \frac{a(m_1,m_2,l)}{a(m_1,m_2)} \;(2i) [\;
 -
 ({\vec S}_{m_1} \times {\vec S}_{l})^-
 -
 ({\vec S}_{m_2} \times {\vec S}_{l})^- \;].
$$
One can verify that
$$
T_{(m_1,m_2)} a(m_1,m_2)\phi(m_1,m_2) = \Phi(m_1,m_2).
$$
further, define
$$
F_{(m_1,m_2)}= T_{(m_1,m_2)}
     \frac{1}{i\theta_1}\frac{1}{i\theta_2}
     \frac{\partial}{\partial m_1}\frac{\partial}{\partial m_2}
$$
we get
$$
(\sum_{m_1<m_2}^{N}F_{(m_1,m_2)}) \mid \psi_2 >  =
       3 \mid \psi_3 >,
$$
on the other hand,(see eq.(\ref{eqqr1}) and eq.(\ref{eqqr2}))
$$
[\; Q_{3,2}^+ = -i \sum_{j<k}^{N} W_{jk}^{(3)}({\vec S}_j \times {\vec S}_k)^-\;]
\mid \psi_2>=\mid \psi_3>,
$$
thus
\begin{equation}
\label{eqq3f}
Q_{3,2}^+ = \frac{1}{3}\sum_{m_1<m_2}^{N}F_{(m_1,m_2)},
\end{equation}
by making comparison the coefficients of $({\vec S}_j \times {\vec S}_k)^-$
of the both sides of eq.(\ref{eqq3f}), it yields
\begin{equation}
\label{soluw3jk}
W_{j,k}^{(3)}=  \frac{2}{3}
\frac{1}{i\theta_1}\frac{1}{i\theta_2}
\{ \;
 \sum_{l\ne j,k}^{N} A(j,k,l)
 [ \frac{1}{A(j,l)} \frac{\partial}{\partial j} -
   \frac{1}{A(k,l)} \frac{\partial}{\partial k} ]
   \frac{\partial}{\partial l}
\; \}.
\end{equation}
with
$$
A(j,k,l)= \left \{
  \begin{array}{ll}
     a(j,k,l)   &  if \;\;j<k<l,  \\
     a(l,j,k)   &  if \;\;l<j<k,   \\
     a(j,l,k)   &  if \;\;j<l<k.   \\
  \end{array}
   \right.
$$

Thus the operator $Q_{3,2}^+$ shifting $\mid \psi_2>$ to $\mid \psi_3>$ is
found.

Making use of the similar analysis, one can obtain the general form of
$Q_{r,r-1}^+$ with the solution of
$$
W_{j,k}^{(r)}  = \frac{2}{r}
\frac{1}{i\theta_1}\frac{1}{i\theta_2}\cdots \frac{1}{i\theta_{r-1}}
\{ \;
 \sum_{l_1,l_2,\cdots,l_{r-2}\ne j,k}^{N}
    A(j,k,l_1,l_2,\cdots,l_{r-2})
$$
\begin{equation}
\label{soluwrjk}
 [\; \frac{1}{A(j,l_1,l_2,\cdots,l_{r-2})} \frac{\partial}{\partial j} -
   \frac{1}{A(k,l_1,l_2,\cdots,l_{r-2})} \frac{\partial}{\partial k} \;] \;
   \frac{\partial}{\partial l_1}
   \frac{\partial}{\partial l_2} \cdots
   \frac{\partial}{\partial l_{r-2}}
\; \}.
\end{equation}

In (b) and (c) one sees that the actions of the partial differential operators
$\frac{1}{i\theta}\frac{\partial}{\partial m}$ and
$\frac{1}{i\theta_1}\frac{1}{i\theta_2}
\frac{\partial}{\partial m_1}\frac{\partial}{\partial m_2}$
are picking up certain terms $a(m)\phi(m)$ and $a(m_1,m_2)\phi(m_1,m_2)$
from $\mid \psi_1>$ and $\mid \psi_2>$ respectively. However, since
$\mid \psi_0>$ has only one single term, we need not to introduce some
partial differential operators to pick it up from $\mid \psi_0>$. That is
the reason why partial differential operators do not emerge in the
coefficients $W_{j,k}^{(1)}$.

An arbitrary state $\mid \psi_r>$ can be obtained by repeated application of
$Q_{r,r-1}^+ $ to the state $\mid \psi_0>$:
\begin{equation}
\label{eqqrto0}
\mid \psi_r>= Q_{r,r-1}^+ Q_{r-1,r-2}^+ \cdots Q_{2,1}^+ Q_{1,0}^+
            \mid \psi_0> ,
\end{equation}
and a caution should be made that when refer to an operator $Q_{r,r-1}^+ $,
it always acts on $\mid \psi_{r-1}>$.

The facts in eq.(\ref{eqqr2}) and eq.(\ref{eqqrto0}) are mostly similar to
those in a hydrogen atom, where the eigenfunctions are expressed in terms of
the radial wavefunction $R_{nl}(r)$ and the angular part wavefunction
$Y_{lm}(\hat{\bf r })$ as
\begin{equation}
\label{atom}
\psi_{nlm}({\bf r})= R_{nl}(r) Y_{lm}(\hat{\bf r }),
\end{equation}
here ${\bf r}$ is the three-dimensional coordinate, $\hat{\bf r }=
r^{-1}{\bf r}$ a unit vector, $Y_{lm}(\hat{\bf r })$ is a spherical
harmonic and
$$
R_{nl}(r)=C_{nl} \;
            _1F_1(-n+l+1;2l+2;\frac{2r}{na})
            (\frac{2r}{na})^l \exp(-\frac{r}{na})
$$
with $C_{nl}$ the normalized constant and $a$ the Bohr radius.

For a given state $\psi_{l+1,l,l}= R_{l+1,l} Y_{ll}$, the unit dipole vector
$ \hat{\bf r } $ will shift the angular part wavefunction as follows:
$$
  { \hat{\bf r } }^\pm Y_{l,\pm l}= \mp (\frac{2(l+1)}{2l+3})^{1/2}
  Y_{l+1,\pm l \pm 1},
$$
where ${ \hat{\bf r } }^+={ \hat{ x } } + { \hat{ y } }$.

To shift the radial part one must introduce the dilatation operator
\begin{equation}
\label{dila}
   D_{n\pm 1, n}^{\pm} =
      \exp[(\frac{i}{\hbar}rp_r+\frac{1}{2}) ln\frac{n}{n\pm 1}],
\end{equation}
with the radial momentum
$p_r=-i \hbar (\frac{\partial}{\partial r}+\frac{1}{r})$, a partial
differential operator. Since
$$
   D_{n\pm 1, n}^{\pm}  \frac{r}{n}= \frac{r}{n\pm 1}  D_{n\pm 1, n}^{\pm},
$$
it leads to
$$
 D_{l+2,l+1}^+ \frac{r}{l+1} R_{l+1,l}(r)=
 -\frac{1}{2}[(2l+3)(2l+4)]^{1/2} R_{l+2,l+1}(r).
$$
As a result
\begin{equation}
\label{shift}
    D_{l+2,l+1}^+ \frac{r}{l+1} { \hat{\bf r } }^+ \psi_{l+1,l,l}({\bf r})
    =a [(l+1)(l+2)]^{1/2}
    \psi_{l+2,l+1,l+1}({\bf r}),
\end{equation}
and a general state $\psi_{l+1,l,l}({\bf r})$ can be generated from the
ground state $\psi_{100}({\bf r})$ by
\begin{equation}
\label{shift100}
    \psi_{l+1,l,l}({\bf r})=
      a^{-l} (l!)^{-1} (l+1)^{-1/2}
      ( D_{l+1,l}^+ \frac{r}{l} ) \cdots
      ( D_{3,2}^+ \frac{r}{2} ) ( D_{2,1}^+ \frac{r}{1} )
       ({ \hat{\bf r } }^+)^l \psi_{100}({\bf r}).
\end{equation}

Eq.(\ref{eqqr2}) and eq.(\ref{shift100}) are quite similar to
eq.(\ref{shift}) and eq.(\ref{eqqrto0}), respectively, based on the following
sense: for the Bethe state $\mid \psi_{r} >$ (see eq.(\ref{eqpsir})), it also
consists of an angular part wavefunction $\phi( m_1,m_2,\cdots,m_r)$ and a
radial part(one-dimensional) wavefunction $a(m_1,m_2,\cdots,m_r)$ that
expresses the phase transition. In the shift operator $Q_{r,r-1}^+$,
the action of $ -i ({\vec S}_j \times {\vec S}_k)^-$ is transforming the
angular part, and the partial differential operaor $W_{jk}^{(r)}$ transforms
the radial part, just like the dilatation operator in eq.(\ref{shift}) dose.
Certainly, there are some differeces between the Hydrogen atom and the
XXX-Heisenberg chain, for the former its wavefunction $\psi_{nlm}$ is
simultaneous eigenstate of the set $\{H_{atom}, L^2, L_z \}$, while
for the latter $\mid \psi_{r} >$ is simultaneous eigenstate of the
set $\{H_{XXX}, S_z \}$, but not an eigenstate of the square of the total
spins operator ${\bf S}^2$.

Eventually, we would like to point out: (i) the shift operator $Q_{1,0}^+ $
can also be obtained from physical consideration the same as that developing
in (b) and (c) (see Appendix A); (ii) From eq.(\ref{eqqr2}), we see that
$Q_{r,r-1}^+ $ are raising operators, which can also be found following the
same idea (see Appendix B). In conclusion, we have established shift
operators for XXX-Heisenberg chain. It is interesting to extent the method to the
Hubbard model\cite{liebwu}\cite{korepin} and the generalized Bethe
ansatz\cite{yang} in subsequent investigations.

{\bf Acknoledgment}

This work was partially supported by the National Natural Science Foundation
of China.


\pagebreak

{\bf APPENDIX A: Clear Physical Picture for Shifting $\mid \psi_0>$ to
$\mid \psi_1>$}

\vspace{1cm}

Because
\begin{equation}
-i ({\vec S}_j \times {\vec S}_k)^- \mid \stackrel{j}{\uparrow}
\stackrel{k}{\uparrow} > = \frac{1}{2}
(\mid \stackrel{j}{\downarrow} \stackrel{k}{\uparrow} > -
\mid \stackrel{j}{\uparrow} \stackrel{k}{\downarrow} > ),
\;\;\;
(S_j^-+S_k^-)\mid \stackrel{j}{\uparrow} \stackrel{k}{\uparrow} >
=
(\mid \stackrel{j}{\downarrow} \stackrel{k}{\uparrow} > +
\mid \stackrel{j}{\uparrow} \stackrel{k}{\downarrow} > ),
\end{equation}
so that
$$
\mid \stackrel{j}{\downarrow} \stackrel{k}{\uparrow} >
=
-i ({\vec S}_j \times {\vec S}_k)^-   +
\frac{1}{2}(S_j^-+S_k^-)
\mid \stackrel{j}{\uparrow} \stackrel{k}{\uparrow} >,
$$
\begin{equation}
\label{spin}
\mid \stackrel{j}{\uparrow} \stackrel{k}{\downarrow} >
=
i ({\vec S}_j \times {\vec S}_k)^-   +
\frac{1}{2}(S_j^-+S_k^-)
\mid \stackrel{j}{\uparrow} \stackrel{k}{\uparrow} >.
\end{equation}

Due to eq.(\ref{spin}), one finds the transformation
$$
\mid \stackrel{1}{\uparrow} \stackrel{2}{\uparrow} \cdots
\stackrel{m}{\uparrow} \cdots \stackrel{N}{\uparrow}  >
  \Longrightarrow
a(m)\phi(m)=
a(m) \mid \stackrel{1}{\uparrow} \stackrel{2}{\uparrow} \cdots
\stackrel{m}{\downarrow} \cdots \stackrel{N}{\uparrow}  >
\;\;\;(m=1,2,\cdots,N),
$$
can be realized by the operator:
\begin{equation}
 T_{0 \rightarrow m}^-= \frac{1}{N-1}a(m) \{ \;
     \sum_{j=1}^{m-1} [\;
     i ({\vec S}_j \times {\vec S}_m)^-  + \frac{1}{2}(S_j^-+S_m^-) \;]
   +
     \sum_{j=m+1}^{N} [\;
    -i ({\vec S}_m \times {\vec S}_j)^-  + \frac{1}{2}(S_m^-+S_j^-) \;]
    \}.
\end{equation}
Define
$$
 F_0=\sum_{m=1}^{N} T_{0 \rightarrow m}^- =
$$
\begin{equation}
 \label{eqfpsi}
 -i \frac{1}{N-1}\sum_{j<k}^{N}[a(j)-a(k)] ({\vec S}_j \times {\vec S}_k)^-
+
 \frac{1}{N-1}\sum_{m=1}^{N}[\frac{N-2}{2}a(m)+
 \frac{1}{2} \sum_{k=1}^{N}a(k) ]S_m^-,
\end{equation}
one gets
\begin{equation}
  \label{eqfpsin}
 F_0 \mid \psi_0 >
=\sum_{m=1}^{N} a(m)\phi(m) = \mid \psi_1 >
\end{equation}

Using
$$
  \sum_{m=1}^{N} a(m)=0,
$$
and
\begin{equation}
 \sum_{m=1}^{N}a(m)S_m^- \mid \psi_0 >=\mid \psi_1 >,
\end{equation}
from eq.(\ref{eqfpsi}) and  eq.(\ref{eqfpsin}) we then have
\begin{equation}
 -i \frac{1}{N-1}\sum_{j<k}^{N}[a(j)-a(k)] ({\vec S}_j \times {\vec S}_k)^-
\mid \psi_0 >= (1-\frac{1}{2}\frac{N-2}{N-1}) \mid \psi_1 >,
\end{equation}
thus
\begin{equation}
\label{eqq1n}
 -i \frac{2}{N}\sum_{j<k}^{N}[a(j)-a(k)] ({\vec S}_j \times {\vec S}_k)^-
\mid \psi_0 >= \mid \psi_1 >,
\end{equation}
on the other hand
\begin{equation}
\label{eqq1}
[\; Q_1^- = -i \sum_{j<k}^{N} W_{jk}^{(1)}({\vec S}_j \times {\vec S}_k)^-\;]
\mid \psi_0>=\mid \psi_1>,
\end{equation}
by comparing the coefficients of $({\vec S}_j \times {\vec S}_k)^-$ of
both side of eq.(\ref{eqq1}) and eq.(\ref{eqq1n}) , it leads to
\begin{equation}
\label{newwjk1}
W_{jk}^{(1)}=\frac{2}{N}(\exp (ij\theta)-\exp (ik\theta)),\;\;
(j,k=1,2,\cdots,N)
\end{equation}
which is nothing but eq.(\ref{eqwjk1}). Thus the physical picture for the
transformation from $\mid \psi_0>$ to $\mid \psi_1>$ is also clear.

\vspace{1cm}

{\bf APPENDIX B: The Lowering Operators}

\vspace{1cm}

Because $i ({\vec S}_j \times {\vec S}_k)^+=S_j^+ S_k^z - S_k^+S_j^z$,
one can calculate that
\begin{equation}
\label{eqsjk+2}
i ({\vec S}_j \times {\vec S}_k)^+ \mid \stackrel{j}{\uparrow}
\stackrel{k}{\downarrow} > = -\frac{1}{2}
\mid \stackrel{j}{\uparrow} \stackrel{k}{\uparrow} > ;
\;\;\;
i ({\vec S}_j \times {\vec S}_k)^+ \mid \stackrel{j}{\downarrow}
\stackrel{k}{\uparrow} > = \frac{1}{2}
\mid \stackrel{j}{\uparrow} \stackrel{k}{\uparrow} > .
\end{equation}
We set the Ansatz:
\begin{equation}
\label{eqqr+3}
[\; Q_{0,1}^- = i \sum_{j<k}^{N} {W'_{jk}}^{(1)}
 ({\vec S}_j \times {\vec S}_k)^+ \;] \;
 \mid \psi_{1}>=\mid \psi_{0}>,
\;\; {W'_{jk}}^{(1)}=-{W'_{kj}}^{(1)},
\end{equation}
we obtain the equation
\begin{equation}
\label{eqwr'1}
\frac{1}{2} \sum_{m=1}^{N} [\;
 (-\sum_{j=1}^{m-1} {W'_{jm}}^{(1)}
 +\sum_{k=m+1}^{N} {W'_{mk}}^{(1)})
 a(m) \;]=1,
\end{equation}
whose solution are
\begin{equation}
\label{soluw'jk}
{W'_{j,k}}^{(2)}= \frac{2}{N(N-1)} \frac{1}{i\theta}
\;[ \frac{1}{a(j)} \frac{\partial}{\partial (m=j)}-
  \frac{1}{a(k)} \frac{\partial}{\partial (m=k)} ].
\end{equation}

Becuase
\begin{equation}
i ({\vec S}_j \times {\vec S}_k)^+ \mid \stackrel{j}{\downarrow}
\stackrel{k}{\downarrow} > = \frac{1}{2}
(\mid \stackrel{j}{\downarrow} \stackrel{k}{\uparrow} > -
\mid \stackrel{j}{\uparrow} \stackrel{k}{\downarrow} > ),
\;\;\;
(S_j^++S_k^+)\mid \stackrel{j}{\downarrow} \stackrel{k}{\downarrow} >
=
(\mid \stackrel{j}{\downarrow} \stackrel{k}{\uparrow} > +
\mid \stackrel{j}{\uparrow} \stackrel{k}{\downarrow} > ),
\end{equation}
so that
$$
\mid \stackrel{j}{\downarrow} \stackrel{k}{\uparrow} >
=
i ({\vec S}_j \times {\vec S}_k)^+   +
\frac{1}{2}(S_j^++S_k^+)
\mid \stackrel{j}{\downarrow} \stackrel{k}{\downarrow} >,
$$
\begin{equation}
\label{spin+}
\mid \stackrel{j}{\uparrow} \stackrel{k}{\downarrow} >
=
-i ({\vec S}_j \times {\vec S}_k)^+   +
\frac{1}{2}(S_j^++S_k^+)
\mid \stackrel{j}{\downarrow} \stackrel{k}{\downarrow} >.
\end{equation}

Therefore, based on eq.(\ref{spin+}) there would be some differences in
achieving shift operators $Q_{r-1,r}^-$ when $r>1$. In the position,
we set the Ansatz:
\begin{equation}
\label{eqqr3}
[\; Q_{r-1,r}^- = i \sum_{j<k}^{N} {W'_{jk}}^{(r)}
 ({\vec S}_j \times {\vec S}_k)^+ \sum_{1}^{N}
 \lambda_j^{(r)} S_j^+ \;] \;
 \mid \psi_{r}>=\mid \psi_{r-1}>,
\;\; {W'_{jk}}^{(r)}=-{W'_{kj}}^{(r)}.
\end{equation}
For instance, when $r=2$, since
$$
  \mid \psi_2 >=\sum_{m_1<m_2} a(m_1,m_2) \mid \cdots
  \stackrel{m_1}{\downarrow} \cdots \stackrel{m_2}{\downarrow} \cdots  >,
$$
$$
  \mid \psi_1 >=\frac{1}{N-1} \sum_{m_1<m_2}[\; a(m_1) \mid \cdots
  \stackrel{m_1}{\downarrow} \cdots \stackrel{m_2}{\uparrow} \cdots  >+
  a(m_2) \mid \cdots
  \stackrel{m_1}{\uparrow} \cdots \stackrel{m_2}{\downarrow} \cdots  >\;],
$$
owing to eq.(\ref{spin+}), one finds the transformation
$$
a(m_1,m_2) \mid \cdots
  \stackrel{m_1}{\downarrow} \cdots \stackrel{m_2}{\downarrow} \cdots  >
  \Longrightarrow
a(m_1) \mid \cdots
  \stackrel{m_1}{\downarrow} \cdots \stackrel{m_2}{\uparrow} \cdots  >
$$
can be realized by the operator:
\begin{equation}
 T_{(m_1,m_2) \rightarrow m_1}^-= \frac{a(m_1)}{a(m_1,m_2)}
      [\;
     i ({\vec S}_{m_1} \times {\vec S}_{m_2})^+  +
     \frac{1}{2}(S_{m_1}^+ + S_{m_2}^+) \;],
\end{equation}
and the transformation
$$
a(m_1,m_2) \mid \cdots
  \stackrel{m_1}{\downarrow} \cdots \stackrel{m_2}{\downarrow} \cdots  >
  \Longrightarrow
a(m_2) \mid \cdots
  \stackrel{m_1}{\uparrow} \cdots \stackrel{m_2}{\downarrow} \cdots  >
$$
is realized by the operator:
\begin{equation}
 T_{(m_1,m_2) \rightarrow m_2}^-= \frac{a(m_2)}{a(m_1,m_2)}
      [\;
    -i ({\vec S}_{m_1} \times {\vec S}_{m_2})^+  +
    \frac{1}{2}(S_{m_1}^+ + S_{m_2}^+) \;].
\end{equation}

Hence,
\begin{equation}
 [\; T_{(m_1,m_2) \rightarrow m_1}^-  +
     T_{(m_1,m_2) \rightarrow m_2}^- \;]
     a(m_1,m_2)\phi (m_1,m_2)=
     a(m_1)\phi (m_1) + a(m_2)\phi (m_2).
\end{equation}
Define
\begin{equation}
 F'_{(m_1,m_2)}=  [\; T_{(m_1,m_2) \rightarrow m_1}^-  +
     T_{(m_1,m_2) \rightarrow m_2}^- \;]
     \frac{1}{i\theta_1}\frac{1}{i\theta_2}
     \frac{\partial}{\partial m_1}\frac{\partial}{\partial m_2},
\end{equation}
so that
$$
(\sum_{m_1<m_2}^{N}F_{(m_1,m_2)}) \mid \psi_2 >  =
       (N-1) \mid \psi_1 >,
$$
on the other hand,
$$
 Q_{1,2}^- = i \sum_{j<k}^{N} {W'_{jk}}^{(2)}
 ({\vec S}_j \times {\vec S}_k)^+ \sum_{1}^{N} \lambda_j^{(2)} S_j^+,
$$
thus
\begin{equation}
\label{eqq12f}
Q_{1,2}^- = \frac{1}{N-1}\sum_{m_1<m_2}^{N}F'_{(m_1,m_2)},
\end{equation}
by making comparison the coefficients of $({\vec S}_j \times {\vec S}_k)^+$
and $S_j^+$ of the both side of eq.(\ref{eqq12f}), it yields
$$
W_{m_1,m_2}^{(2)}=  \frac{1}{N-1}  \frac{a(m_1)-a(m_2)}{a(m_1,m_2)}
\frac{1}{i\theta_1}\frac{1}{i\theta_2}
 \frac{\partial}{\partial m_1} \frac{\partial}{\partial m_2},
 \;\; (m_1< m_2),
$$
\begin{equation}
\label{soluw2'}
\lambda_j^{(2)} = \frac{1}{N-1} \frac{1}{i\theta_1}\frac{1}{i\theta_2}
[\; \sum_{m=1}^{j-1} \frac{a(m)+a(j)}{a(m,j)}
 \frac{\partial}{\partial m} \frac{\partial}{\partial j}
 +
   \sum_{k=j+1}^{N} \frac{a(j)+a(k)}{a(j,k)}
 \frac{\partial}{\partial j} \frac{\partial}{\partial k} \;].
\end{equation}

Lowering operators for $r>3$ can be obtained in the same way.

      }
\end{document}